\documentclass[aip,floats,superscriptaddress]{revtex4}

\usepackage{graphicx}
\usepackage{epstopdf}
\usepackage{amssymb}
\usepackage{amsmath,bm}
\usepackage{psfrag}
\usepackage{epsfig}
\usepackage{float}
\usepackage{bm}

\begin{document}

\title{Electromigration-guided composition patterns in thin alloy films: a computational study}

\author{Mikhail Khenner\footnote{Corresponding
author. E-mail: mikhail.khenner@wku.edu.}}
\affiliation{Department of Mathematics, Western Kentucky University, Bowling Green, KY 42101, USA}
\affiliation{Applied Physics Institute, Western Kentucky University, Bowling Green, KY 42101, USA}

\begin{abstract}
\noindent
Via computation of a continuum dynamical model of the diffusion and electromigration, this paper demonstrates the feasibility 
of guiding the formation of the stripe composition patterns in the thin surface layers of the crystal alloy films.
By employing the systematic parametric computational analysis it is revealed how such properties of the pattern as the aerial number density 
of the stripes and the stripe in-plane orientation are influenced by the major physical factors that are not limited to the electric field 
strength and its direction angle in the plane, but also include a number of parameters that originate in the anisotropy of diffusion in the particular 
crystallographically-oriented surface layer. By following the insights from this analysis the real patterns 
hopefully can be created in a dedicated experiment.

\end{abstract}

\date{\today}
\maketitle


\section{Introduction}
\label{Intro}

Surface electromigration \cite{YTRW,TCW} is widely used in thin solid films research for, for example, forming nanocontacts \cite{VFDMSKBM,GSPBT}, guiding 
motion of monoatomic steps and 
islands on crystal surfaces \cite{YTRW,TCW,CMECPL}, and forming facets on crystal surfaces \cite{ZYDZ}.
Modeling studies of the effects produced by surface electromigration in thin, monocomponent single-crystal films include 
surface step bunching \cite{St,DDF,CPM,OPL}, surface faceting \cite{KD,SK,DS,BMOPL}, 
harnessing of surface morphology instabilities caused by stress and wetting of a substrate \cite{THP,M1,TGM1,O,K}, asperity contact evolution in microelectromechanical systems switches \cite{KSCY}, 
control of surface roughness \cite{DM} and islands or nanowire morphology \cite{CMECPL,KKHV,DasM,DKM,KDDM}, 
and driving adsorbate transport on graphene \cite{SV}.

It seems plausible that electromigration should be capable of driving the formation of ordered 
composition patterns in a thin, flat surface layer of a substitutional, single-crystal, metallic alloy thin film.\footnote{Morphological evolution may be suppressed by the same
external electric field that drives the proposed formation of a composition pattern \cite{DM,K}.}
However, we are not aware of any studies of such pattern-forming system.
Traditional studies of the electromigration effects in alloys are limited to the bulk, polycrystal conducting lines and related microelectronic 
reliability issues \cite{SKA,PAT,DVAG,YT}.
A system that seems most amenable for displaying the surface electromigration-driven composition patterning effects is the A$_{1-x}$B$_x$ surface alloy \cite{SurfAlloying1,SurfAlloying2}, 
where intermixing of A and B atoms occurs in the first few atomic layers \cite{TersoffAlloy}.
Another system could be simply a continuous, single-crystal, A$_{1-x}$B$_x$ ultrathin film on a thick substrate.
Formation of ordered surface composition patterns is important for heterogeneous catalysis, 
corrosion, lubrication and adhesion. Also the electric, magnetic, plasmonic, and photovoltaic properties of a surface can
be strongly influenced by the composition of the near surface region. But to our knowledge, 
no experiment or model study of electromigration in alloys was published where the compositional variations are presented, 
and the means of guiding these variations have not been proposed.\footnote{One exception to this is the one-dimensional model in Ref. \cite{KB}, where surface electromigration 
in a binary alloy film is coupled to evolution of the surface morphology.} 

In this paper we employ a computational study of a surface electromigration model to demonstrate that 
the external electric fields are capable of creating stripe patterns of the atomic composition (concentration) in thin films. 
We show that many properties of these patterns, such as the pattern uniformity, its (mis)alignment to the direction of the applied electric field, 
and the aerial number density of the stripes, depend on the electric field strength and on the physical parameters entering the anisotropic diffusivity tensor for a particular low-index surface that bounds 
the thin diffusive layer from a vacuum (or vapor) side.  Thus by selecting a special combination of these parameters in experiment 
(through the choice of the alloy, low-index surface, direction of the applied electric field, etc.) it should be possible to guide the stripe pattern.

\section{The Model}
\label{Model}

\begin{figure}[H]
\vspace{-0.2cm}
\centering
\includegraphics[width=4.0in]{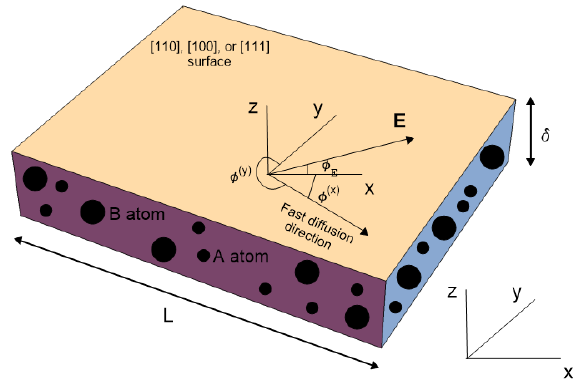}
\vspace{-0.15cm}
\caption{Schematic representation of a conducting binary alloy film subjected to an external electric field $\bm{E}$.
}
\label{Fig1}
\end{figure}

We consider a compositionally nonuniform binary metal layer (a substitutional binary alloy) of a constant thickness $\delta$, annealed at high temperature.
The layer may represent either the surface layer of a surface alloy film,
or the ultrathin alloy film on a thick conductive substrate. We refer to this setup simply as film.
Two atomic components of the film are denoted as A and B. 
The electric potential difference (voltage) $\Delta V$ is applied to the opposite edges of the substrate, which results in
the electric field $\bm{E}=\left(E_0\cos{\phi_E},E_0\sin{\phi_E}\right),\; E_0=\Delta V/L$ (Fig. \ref{Fig1}). 
Here $L$ is the width of the substrate between the anode and the cathode, and also the horizontal film dimension, and $\phi_E$ the angle that the electric 
field vector makes with the $x$-axis of the Cartesian reference frame (the direction angle). $x$ and $y$ axes are in the film/substrate plane, and the $z$ axis is 
perpendicular to that plane, and it points into the vapor or vacuum above the film.
The electric current through the film activates electromigration.
The film thickness $\delta$ is assumed small, i.e. on the order of several atomic diameters \cite{TersoffAlloy}. 
For computations we chose $\delta$ equal to two diameters of a Pd atom (Table \ref{T1}).
Due to small film thickness and to the planar geometry of the electric field, the diffusion and electromigration mass flows along the $z$-axis are negligible 
and thus they are not considered by the model. Computations in Sec. \ref{CompositionEvolve} will be done for CuPd alloy; this alloy is often used in 
electromigration experiments \cite{PV}.

Let $\nu_A$ and $\nu_B$ be the number of atoms of the components A and B per substrate area.
Then the dimensionless concentrations, or the local composition fractions, are defined as $C_A=\nu_A/\nu$ and $C_B=\nu_B/\nu$, where $\nu=\nu_A+\nu_B$. Clearly,
\begin{equation}
C_A(x,y,t)+C_B(x,y,t)=1,
\label{sumC}
\end{equation}
where $t$ is the time. Physically, this condition stems from the negligible vacancy concentration in a substitutional alloy. 
In the remainder of this section we construct a diffusion/electromigration model 
that enables to follow the spatio-temporal evolution of $C_B$. Notice that this is sufficient, since due to Eq. (\ref{sumC}) the evolution of 
$C_A$ follows trivially from equation $\partial C_A/\partial t = - \partial C_B/\partial t$ (thus if $C_B$ increases locally, i.e. at some point $(x,y)$ in the film, 
then $C_A$ decreases at that point with the same rate). 

%

The diffusion equation reads  
\begin{equation}
\frac{\partial C_B}{\partial t} =   -\frac{\Omega}{\delta} \bm{\nabla} \cdot \bm{J}_B, \label{C-eq}
\end{equation}
where $\Omega$ is the atomic volume and $\bm{\nabla}=(\partial_x, \partial_y)$. Since the 
mass transport is driven by the gradient of the chemical potential $\mu_B$ and by electromigration, the flux $\bm{J}_B$ is given by
\begin{equation}
\bm{J}_B = -\frac{\nu_B}{kT}\bm{D}_B\cdot \left(\bm{\nabla} \mu_B+q\bm{E}\right) = -\frac{\nu}{kT}C_B\bm{D}_B\cdot \left(\bm{\nabla} \mu_B+q\bm{E}\right),
\label{J-eq}
\end{equation}
where $\bm{D}_B$ is the diffusion tensor, $kT$ the Boltzmann's factor, and $q>0$ the effective charge.

The chemical potential in Eq. (\ref{J-eq}) is the sum of the contributions due to alloy thermodynamics and the compositional stress \cite{ZVD,SVT,RSN}:
\begin{equation}
\mu_B = \Omega \left[\frac {1-C_B}{\delta}\frac{\partial \gamma}{\partial C_B} + \tau \eta_0 \left(1-C_B\right) C_B -\epsilon \bm{\nabla}^2 C_B\right],
\label{base_eq4}
\end{equation}
where $\gamma$ is the free energy, 
$\epsilon$ the Cahn-Hilliard gradient energy coefficient,
$\eta_0>0$ the effective solute expansion coefficient, and $\tau=\pm 1$. The effective solute expansion coefficient measures the relative size difference of A and B atoms and
quantifies the linear lattice strain due to increasing the concentration of B. $\eta_0>0$ corresponds to large B atoms. Choosing $\tau=1$ or $-1$  depends on whether
the compositional stress is compressive or tensile. When the stress is tensile and $\eta_0>0$, the destabilization of the base uniform composition state is expected \cite{SVT}. 
The linear stability analysis (LSA, see below) shows that for the average typical film composition $A_{0.7}B_{0.3}$ the destabilization occurs at $\tau=-1$. 
Therefore this value corresponds to the tensile stress, and it is chosen for the remainder of this section and for computations in Sec. \ref{CompositionEvolve}.

The free energy in Eq. (\ref{base_eq4}) is written as \cite{RSN,LuKim}:
\begin{equation}
\gamma=\gamma_A\left(1-C_B\right)+\gamma_B C_B +k T \nu \left[\left(1-C_B\right)\ln \left(1-C_B\right)+C_B\ln C_B+ H\left(1-C_B\right)C_B\right],
\label{gamma}
\end{equation}
where $\gamma_A$ and $\gamma_B$ are the energies of pure A and B on the substrate surface, $kT\nu$ is the alloy entropy, and the dimensionless number 
$H=\alpha_{int}/k T\nu$ measures the bond strength relative to the thermal energy $k T$. Here $\alpha_{int}$ is the enthalpy.
The first two terms are needed since the film thickness is on the order of a few atomic diameters, and the terms in the bracket are the regular solution model.
When $H>2$, the graph of $\gamma\left(C_B\right)$ is a double-well curve, which results in spinodal decomposition. 
For simulations we chose $\alpha_{int}$ value (see Table \ref{T1}) such that $H=1.7$ and the $\gamma$-curve is convex, thus spinodal instability is absent and the only source of instability  
is the compositional stress, again refer to the LSA below.\footnote{Experimental conditions are possible when the spinodal instability is present 
alongside with the compositional stress instability. Computations of compositional patterning in such situation in the absence of electromigration can be found in Ref. \cite{KH1}.
Generally, phase separation is enhanced, i.e. it develops faster and the resultant composition differences are larger, when two instabilities act simultaneously.}
 
Finally, the diffusivity, $\mathbf{D}_B$, is given by a transversely isotropic diffusion tensor 
\begin{equation}
\mathbf{D}_B = 
\begin{pmatrix}
D_B^{(xx)} & 0\\
0 & D_B^{(yy)}
\end{pmatrix}
=
\begin{pmatrix}
D_{B,min}^{(xx)}f(\phi^{(x)}) & 0\\
0 & D_{B,min}^{(yy)}f(\phi^{(y)})
\end{pmatrix},
\label{DiffTensor}
\end{equation}
where $f(\phi^{(x)})= 1+\beta\cos^2{[m\phi^{(x)}]}$, $f(\phi^{(y)})= 
1+\beta\cos^2{[m\phi^{(y)}]}$ are the surface diffusional anisotropy functions for face-centered cubic (fcc) crystals \cite{DM,DM1}. $\beta\ge 0$ is the strength of \emph{crystallographic} anisotropy 
and $\phi^{(\alpha)}$ ($\alpha=x,y$) are the misorientation angles formed between the $\alpha$-axis and the fast surface diffusion direction. Note that
$f(\phi^{(x)}),\ f(\phi^{(y)})\ge 1$ (are positive).
The integer $m=1,2,3$ is determined by the crystallographic orientation of the film surface.
For $m=1$ ([110] surface): $\phi^{(y)} = \pi/2+\phi^{(x)}$, $0\le \phi^{(x)}\le \pi/2$; for $m=2$ ([100] surface): $\phi^{(y)} = \phi^{(x)}$, $0\le \phi^{(x)}\le \pi/4$; 
for $m=3$ ([111] surface): $\phi^{(y)} = \pi/6+\phi^{(x)}$, $0\le \phi^{(x)}\le \pi/6$ \cite{DM}. See Refs. \cite{DM,KH1} for plots of $f(\phi^{(x)})$ and $f(\phi^{(y)})$.

Combining equations, using the typical thickness $h=0.1L$ of the as-deposited surface alloy film as the length scale and $kTh^2\delta^2/\Omega^2 \nu D_{B,min}^{(xx)} \gamma_B$ 
as the time scale, yields the dimensionless PDE for $C_B(x,y,t)$:
\begin{equation}
\frac{\partial C_B}{\partial t} = \bm{\nabla}\cdot\left[C_B\bm{\nabla}_{\Lambda B}\left(\left(1-C_B\right) 
\frac{\partial \gamma}{\partial C_B}-G^{(CH)}\bm{\nabla}^2 C_B- S C_B \left(1-C_B\right) \right)\right]+\bm{\nabla}\cdot\left[C_B\left(1-C_B\right) \bm{F}\right],  
\label{nondim_C_eq2_only_final_2D1}
\end{equation}
where 
\begin{equation}
\gamma = \Gamma\left(1-C_B\right) + C_B  + N\left[\left(1-C_B\right) \ln \left(1-C_B\right)+C_B\ln C_B+H \left(1-C_B\right) C_B\right].
\label{nondim_gamma}
\end{equation}
The dimensionless parameters are the ratio of the pure energies $\Gamma=\gamma_A/\gamma_B$, the entropy $N=k T\nu/\gamma_B$, the enthalpy $H=\alpha_{int}/kT \nu$,
the Cahn-Hilliard gradient energy coefficient $G^{(CH)}=\epsilon \delta/h^2\gamma_B$,
the solute expansion coefficient $S= \eta_0 \delta/\gamma_B$, and the electric field strength $F_e=q \delta h \Delta V/L \Omega \gamma_B$ that enters the effective 
``anisotropic" electric field vector 
$\bm{F}=(F_e f(\phi^{(x)})\cos{\phi_E}, F_e \Lambda_B f(\phi^{(y)})\sin{\phi_E})$.
Also $\bm{\nabla}_{\Lambda B}=(f(\phi^{(x)})\partial_x,\Lambda_B f(\phi^{(y)})\partial_y)$ is the anisotropic gradient operator,
where $\Lambda_B=D_{B,min}^{(yy)}/D_{B,min}^{(xx)}$ is the dimensionless ratio of the amplitudes of the diagonal components of the anisotropic diffusivity tensor. 
We will call $\Lambda_B$ the \emph{diffusional} anisotropy. At $\Lambda_B=1$ the diffusional anisotropy is zero.\footnote{Of course, the separation into crystallographic and 
diffusional anisotropy is only for convenience of discussion. Both anisotropies stem from the anisotropy of diffusion in the surface layer of a single-crystal film, 
see Eq. (\ref{DiffTensor}).} $\Lambda_B$ may be as large as 1000, and as small as 0.001 if the atomic rows happen to run along one of the two 
coordinate axes, say the $x$-axis (and correspondingly, the $y$-axis is orthogonal to the rows). 
This is because the diffusion across the rows requires an activation energy that is at least three times
larger than the activation energy for the diffusion along the rows \cite{ES,MG}.

It is clear that the second term in Eq. (\ref{nondim_C_eq2_only_final_2D1}) describes the forced mass transport due to electromigration effect.
Including $\Lambda_B,\ m$ and the angles $\phi^{(x)},\ \phi_E$, there is ten parameters (nine if $\Gamma=1$), since $\phi^{(y)}$ is defined through 
$\phi^{(x)}$.\footnote{Notice that $\Gamma=1$ when $\gamma_A=\gamma_B$ (Table \ref{T1}); 
in this situation the number of dimensionless parameters can be reduced from nine to eight, since $N$ can be made equal to one via replacing $\gamma_B$ by $k T\nu$ 
in the definitions of the remaining parameters.}

\emph{Remark 1.}\; Setting $\beta=0$ amounts to consideration of a generic film surface with the diffusional anisotropy, since  in this case the 
crystallographic parameters $m$, $\phi^{(x)}$ and $\phi^{(y)}$ are irrelevant.
Note that $f(\phi^{(x)})=f(\phi^{(y)})=1$ and $\bm{\nabla}_{\Lambda B}=(\partial_x,\Lambda_B \partial_y)$. It follows that if $\Lambda_B=1$ in addition to $\beta=0$,
then the diffusion tensor is isotropic, with identical diagonal components ($\mathbf{D}_B=D_B\bm{I}$, where $D_B$ is the diffusivity and $\bm{I}$ the identity
tensor), and also $\bm{\nabla}_{\Lambda B}=\bm{\nabla}$. In this case there is complete isotropy.

\emph{Remark 2.}\; When the crystallographic anisotropy strength $\beta > 0$, the 
anisotropic gradient operator $\bm{\nabla}_{\Lambda B}\neq \bm{\nabla}$ even at zero diffusional anisotropy, due to intrinsic reconstruction of a particular 
low-index crystal surface. The overall anisotropy effect is enhanced if in addition to crystallographic anisotropy
the diffusional anisotropy is also present. 

\emph{Remark 3.}\; As derived, the electromigration term should be $\bm{\nabla}\cdot\left[C_B \bm{F}\right]$.
Based on many simulations we realized that multiplication by a "volume exclusion" factor $1-C_B$ is necessary in order to ensure stable computation. 
The factor is due to the fact that the diffusion flow can only pass through the available vacant sites (a finite occupancy effect).
We later found that this factor can be transparently derived by scaling to continuum limit the discrete-space, continuous-time equation for the density of diffusing, size-zero particles on a 
lattice \cite{P}. 
In the field of materials modeling, this factor can be seen, for example, in the forcing terms of the diffusion equations in Refs. \cite{CTT,KDB,KK}. 

\emph{Remark 4.}\; Eq. (\ref{nondim_C_eq2_only_final_2D1}) has been developed in the general spirit of a model for the dynamics of a 
composition of a binary monolayer, developed by Z. Suo's group; see Ref. \cite{LuKim} and the references therein. 

The standard LSA of Eq. (\ref{nondim_C_eq2_only_final_2D1}) about composition $C_B=C_{B0}$ yields the complex perturbation growth rate 
$\omega\left(k_1,k_2\right)=\omega_r\left(k_1,k_2\right)+i\ \omega_i\left(k_1,k_2\right)$,
where $k_1$ and $k_2$ are the perturbation wavenumbers in the $x$ and $y$ directions, and 
\begin{eqnarray}
\omega_r\left(k_1,k_2\right)&=&C_{B0}\left[k_1^2 W - G^{(CH)}\left(k_1^4+k_1^2k_2^2\right)\right]f(\phi^{(x)})+
C_{B0}\Lambda_B\left[k_2^2 W - G^{(CH)}\left(k_2^4+k_1^2k_2^2\right)\right]f(\phi^{(y)}), \label{omr} \\
\omega_i\left(k_1,k_2\right)&=&F_e\left(1-2C_{B0}\right)\left[k_1f(\phi^{(x)})\cos{\phi_E}+\Lambda_B k_2f(\phi^{(y)})\sin{\phi_E}\right], \label{omi}
\end{eqnarray}
with
\begin{equation}
W = 1-\Gamma+S\left(1-2C_{B0}\right)+N\left[H\left(3-4C_{B0}\right)+\frac{C_{B0}\ln\left[C_{B0}/\left(1-C_{B0}\right)\right]-1}{C_{B0}}\right].
\end{equation}
At $C_{B0}=0.3$ (the choice for the computation in Sec. \ref{CompositionEvolve}), the last two terms in $W$ evaluate to -0.0033 at the parameters from Table \ref{T1}, 
and $1-\Gamma=0$ since $\Gamma=1$ due to $\gamma_A=\gamma_B$ for CuPd alloy; thus $W=-0.0033+0.4S$, and at $S>0.0082$ the positive $k_1^2 W$ and $k_2^2 W$ terms in $\omega_r$
contribute to growth of small wavenumbers, while the negative terms proportional to $G^{(CH)}$ contribute to the decay of large wavenumbers. Note that value of $S$ that 
corresponds to Table \ref{T1} parameters is 0.015. Thus the linear instability displayed by $\omega_r\left(k_1,k_2\right)$ has the long-wavelength character.
The imaginary part of the growth rate, $\omega_i\left(k_1,k_2\right)$ is not zero only when the electric field parameter $F_e\neq 0$. Note that $\omega_i$ does not depend on the
compositional stress $S$ or on alloy thermodynamic parameters $\Gamma,\ H$ and $N$, and also that the anisotropies affect values of both $\omega_r$ and $\omega_i$.
In the presence of the electromigration a perturbation drifts in the $xy$-plane with the speed 
$|\omega_i\left(k_1,k_2\right)|/\sqrt{k_1^2+k_2^2}$ (a traveling wave). This drift in the nonlinear regime may influence the pattern formation.\footnote{The electromigration-driven
surface waves were analyzed by Bradley in a model of a morphology evolution for a single-component film  \cite{B1,B2,B3}.} 
Note that the compositional stress contribution in $W$, and the drift, both vanish at $C_{B0}=1/2$, i.e. at the perfectly balanced average layer composition, 50\%A/50\%B.

\begin{figure}[H]
\vspace{-0.2cm}
\centering
\includegraphics[width=5.5in]{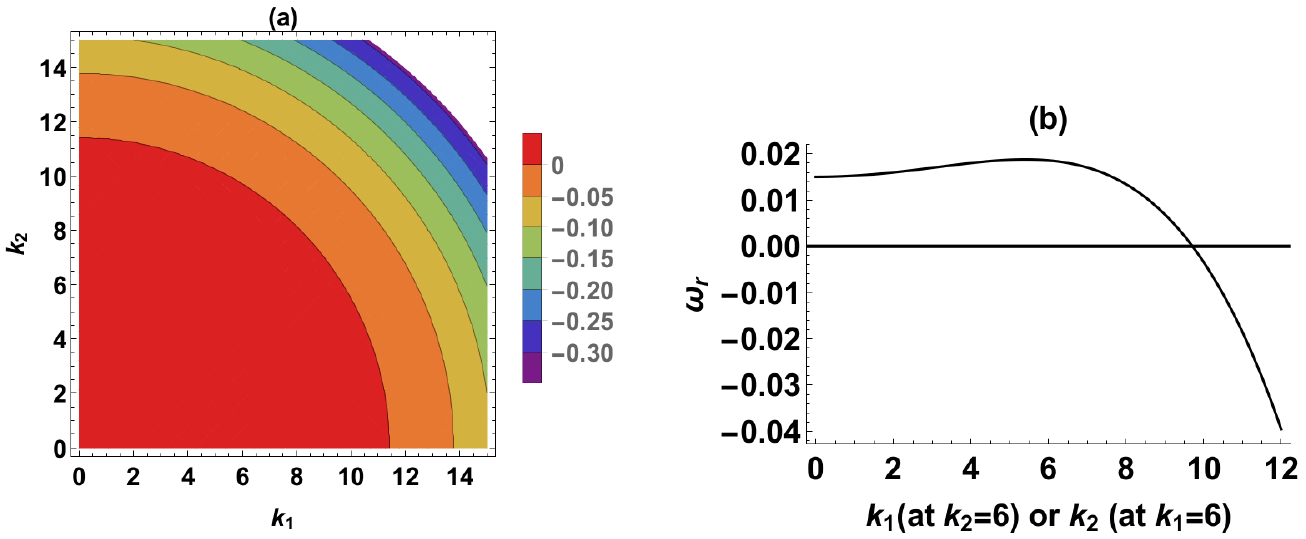}
\vspace{-0.15cm}
\caption{(a): Contour plot of $\omega_r\left(k_1,k_2\right)$ at $\beta=0$ (a generic film surface) and $\Lambda_B=1$. 
Other dimensionless parameters correspond to physical values in Table \ref{T1}. (b): 1D sections of the plot in (a).
}
\label{Fig2new}
\end{figure}

Fig. \ref{Fig2new} shows the plots of $\omega_r$. The maximum value $\omega_r^{max}=0.019$ occurs at $k_1=k_2=6\equiv k_{max}$. When the parameters $\beta$, $\Lambda_B$, $m$, and $\phi^{(x)}$ 
change within the intervals of interest in Sec. \ref{CompositionEvolve}, these plots experience only minor changes, and thus $k_{max}=6$ value changes 
insignificantly. This value sets the size of the (square) computational domain, $\ell=7\lambda_{max}=7(2\pi/k_{max})=7.33$.
\begin{table}[!ht]
\centering
{\scriptsize 
\begin{tabular}
{|c|c|}

\hline
				 
			\rule[-2mm]{0mm}{6mm} \textbf{Physical parameter}	 & \textbf{Typical value}  \\
			\hline
                        \hline
			\rule[-2mm]{0mm}{6mm} $L$ & $5\times 10^{-5}$ cm (500 nm)\\
			\hline
                        \rule[-2mm]{0mm}{6mm} $\delta$ & $6.76\times 10^{-8}$ cm  (0.676 nm) \\
			\hline
				\rule[-2mm]{0mm}{6mm} $\nu$ & $0.5\times 10^{14}$ cm$^{-2}$ \cite{ZVD,LuKim} \\
			\hline
				\rule[-2mm]{0mm}{6mm} $\Omega$ & $2\times 10^{-23}$ cm$^3$  \\
                        \hline
			\rule[-2mm]{0mm}{6mm} $T$ & 923 K \\
			\hline
			    \rule[-2mm]{0mm}{6mm} $q$ & $5\times 10^{-12}$  statC \cite{YTRW} \\
			\hline
			    \rule[-2mm]{0mm}{6mm} $\Delta V$ & $0.03$V    \\	
			\hline
			\rule[-2mm]{0mm}{6mm} $\alpha_{int}$ & $11$  erg$/$cm$^2$  \\ 
			\hline
			    \rule[-2mm]{0mm}{6mm} $\gamma_A$,\; $\gamma_B$ & $2.2\times 10^3$  erg$/$cm$^2$ \cite{VRSK} \\ 
				\hline
			    \rule[-2mm]{0mm}{6mm} $\epsilon$ & $1.2\times 10^{-5}$ erg$/$cm  \cite{Hoyt} \\
			\hline
			\rule[-2mm]{0mm}{6mm} $\eta_0$ & $4.9\times 10^{8}$ erg$/$cm$^3$  \cite{SVT} \\
			\hline

\end{tabular}}
\caption[\quad Physical parameters]{Physical parameters. $\delta=4R_{Pd}, \Omega=4\pi R_{Pd}^3/3, \gamma_A$ and $\gamma_B$ correspond to fcc CuPd alloy, with larger Pd atoms 
taken as the impurity $B$ atoms in the model; in other words, Pd is the solute and Cu is the solvent. $R_{Pd}=0.169$ nm is the calculated radius of a Pd atom \cite{CRR}.
$\eta_0 = 2 Y a^2 p (1+\nu^{(p)})/(1-\nu^{(p)})$, where $Y=\left(Y_{Pd}+Y_{Cu}\right)/2$, $\nu^{(p)}=\left(\nu^{(p)}_{Pd}+\nu^{(p)}_{Cu}\right)/2$, and 
$a=2\left(a_{Pd}-a_{Cu}\right)/\left(a_{Pd}+a_{Cu}\right)$. $Y_{Cu}=1.3\times 10^{12}$erg$/$cm$^3$, $\nu^{(p)}_{Cu}=0.35$, $a_{Cu}=3.6\times 10^{-8}$cm are the Young's modulus, Poisson's ratio, and lattice constant of
Cu, $Y_{Pd}=1.21\times 10^{12}$erg$/$cm$^3$, $\nu^{(p)}_{Pd}=0.39$, $a_{Pd}=3.86\times 10^{-8}$cm are the Young's modulus, Poisson's ratio, and lattice constant of
Pd, and $0< p=0.007< \nu^{(p)}/(1+\nu^{(p)})=0.27$ is the dimensionless residual stress due to the composition variations. 
Note that $E_0$ that corresponds to $\Delta V$ and $L$ in the Table is 600 V/cm, and the dimensionless strength of the electric field, $F_e$, is proportional to
$E_0$. Thus values of $\Delta V$ and $L$ in the Table can be replaced by, say, $\Delta V=30$V and $L=0.05$cm without changing $E_0$ and $F_e$ values. 
This means that the stripe patterns computed in Sec. \ref{CompositionEvolve} can be magnified to fill larger film area.
}
\label{T1}
\end{table}
\section{Formation and evolution of composition patterns}
\label{CompositionEvolve}

In this section we present the results of computations with Eq. (\ref{nondim_C_eq2_only_final_2D1}). 
In all computations we use the initial condition $C_B(x,y,0)=0.3+\xi(x,y)$, where $\xi(x,y)$ is a random deviation from the mean value 0.3 with the 
maximum amplitude 0.08 at a point $(x,y)$ (Fig. \ref{Fig2}(a)).
We employ the method of lines framework with a pseudospectral spatial discretization on 140x140 rectangular grid and a stiff 
ODE solver in time. 
Periodic conditions are imposed on the boundary of a square computational domain. 
We fix the physical parameters to their values in Table \ref{T1}, and systematically vary the direction angle of the electric field $\phi_E$, the diffusional 
anisotropy $\Lambda_B$ (both at the zero crystallographic anisotropy), and the crystallographic orientation of the surface $m$ 
(at the fixed non-zero strength of the crystallographic anisotropy, $\beta=5$ \cite{DM}, at zero diffusional anisotropy, and at fixed $\phi_E$). 
This is in order to transparently separate, and thus understand, the effects of both anisotropies and of the electric field direction angle. 
Also, for [110] surface ($m=1$), we compute the ``full" case, i.e. at both anisotropies present and with the fixed direction angle of the electric field, and varying the 
misorientation ``diffusion" angle $\phi^{(x)}$. And finally, we fix $\phi^{(x)}$ in the full case and study the effect of the electric field strength by moderately varying applied voltage $\Delta V$.
Note that by the choice of $\Delta V$ the electric field parameter $F_e$ is rather small, so that forced mass transport due to electromigration competes with the natural diffusion.

\subsection{Zero electric field $(F_e=0)$}
\label{ZeroEfield}

\begin{figure}[H]
\vspace{-0.2cm}
\centering
\includegraphics[width=5.0in]{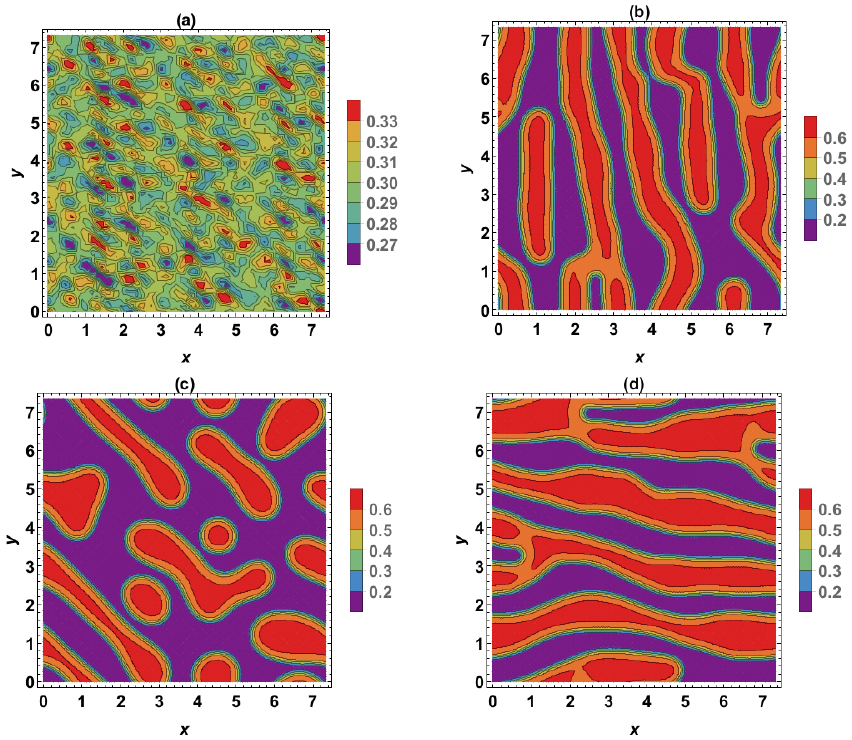}
\vspace{-0.15cm}
\caption{(a): Example of the initial condition, $C_B(x,y,0)$. (b-d): Composition patterns at the late time, here $t=5000$, 
for $\Lambda_B=0.1,\ 1,\ 10$, respectively, in the absence of the electric field $\left(F_e=0\right)$. $\beta=0$ (a generic film surface).
See Sec. \ref{ZeroEfield} for discussion. 
}
\label{Fig2}
\end{figure}

When the external electric field is off, the composition patterns in the thin film are disordered. Fig. \ref{Fig2}(b-d) shows as the example the late time composition 
patterns at $\beta=0$ and $\Lambda_B=0.1,\ 1,\ 10$. In this case 
different patterns arise solely due to diffusional anisotropy, see Remarks 1 and 2 in Sec. \ref{Model}. As expected, at $\Lambda_B=1$ (i.e., when the diffusion strengths in the 
$x$ and $y$ direction are the same, $D_{B,min}^{(yy)}=D_{B,min}^{(xx)}$), $\bm{\nabla}_{\Lambda B}=\bm{\nabla}$ and the spatial rates of change 
in the $x$ and $y$ directions are given by $\partial/\partial x$ and $\partial/\partial y$, thus the spatial variations in the pattern are similar in both directions.
At $\Lambda_B=0.1$, the rate of change in the $y$-direction is roughly 10 times smaller than in the $x$-direction,
thus the pattern looks more uniform in the $y$-direction than in the $x$-direction, i.e. the changes in the form of the alternation of small and large $C_B$ 
occur primarily in the $x$-direction. The overall effect is the pattern that consists of a worm-like, often ``zipped" (forked) bands of alternating large/small $C_B$ content 
(red/purple colored) running in the $y$ direction. At $\Lambda_B=10$ the situation is reversed. Note that the final dimensionless time $t=5000$ in Fig. \ref{Fig2}(b-d) 
corresponds to the physical time of 28 min, 
assuming the typical surface diffusivity value $D_{B,min}^{(xx)}=10^{-9}$cm$^2/$s at $T=923$K \cite{AKR}.

\subsection{Non-zero electric field}
\label{NonZeroEfield}

\subsubsection{Non-zero crystallographic anisotropy, zero diffusional anisotropy: $\beta=5,\ \Lambda_B=1,\ \phi^{(x)}=\pi/6$}
\label{NonZeroEfield_phiE_m_Study}

In this subsection we summarize composition patterns in the films of three chosen crystallographic orientations as a function of 
the direction angle of the electric field, $\phi_E$. 

\begin{enumerate}

\item $\phi_E=0$ or $\pi$.\\
These electric field directions are not equivalent when there is a coupling to the crystallographic anisotropy, as follows from the definition of the effective electric field 
vector $\bm{F}$, see Sec. \ref{Model}. In [110]-oriented film ($m=1$), both electric field directions result in the perfect stripe pattern whose axis is parallel 
to the $x$ axis (Fig. \ref{Fig3}(a)).  This is also the case at $\phi_E=0$ and [100]-oriented film ($m=2$). At all other combinations of two  $\phi_E$ values and 
three $m$ values there still is the orientation effect of the electric field, i.e. the field tries to align the pattern axis with the $x$-axis, however,
the stripes do not emerge. At any time during the evolution, there emerge instead the multiple-connected domains, see Fig. \ref{Fig3}(b).
\begin{figure}[H]
\vspace{-0.2cm}
\centering
\includegraphics[width=5.0in]{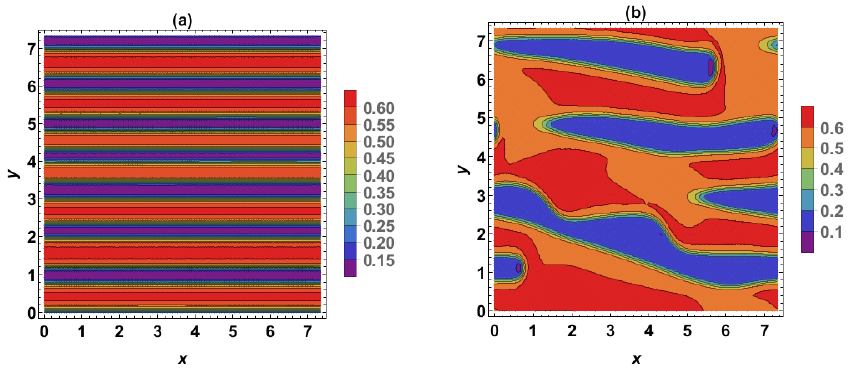}
\vspace{-0.15cm}
\caption{Examples of composition patterns. $C_B(x,y,1000)$ at $\phi_E=\pi$ and (a): $m=1$, (b): $m=3$. See Sec. \ref{NonZeroEfield_phiE_m_Study}, item 1 for discussion.
}
\label{Fig3}
\end{figure}

\item $\phi_E=\pi/2$ or $-\pi/2$.\\
This situation is similar to item 1 in this list. There is either the vertical stripes (Fig. \ref{Fig4}), or the multiple-connected domains loosely aligned with the $y$-axis.
Notice that the aerial number density of stripes, when checked at the same time, varies with the crystallographic orientation.
\begin{figure}[H]
\vspace{-0.2cm}
\centering
\includegraphics[width=5.0in]{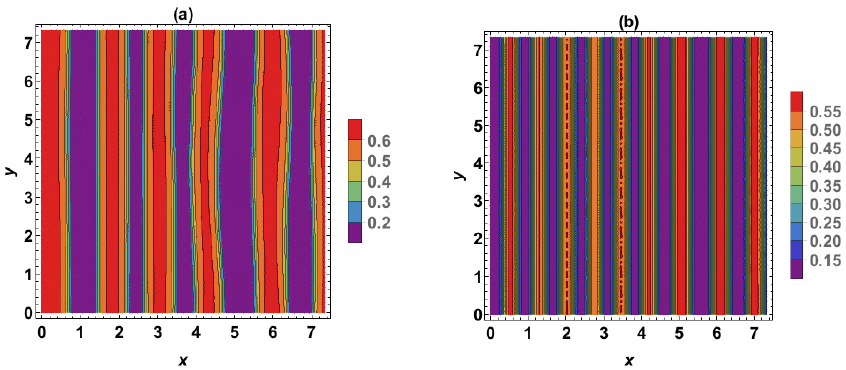}
\vspace{-0.15cm}
\caption{Examples of composition patterns. $C_B(x,y,1000)$ at $\phi_E=\pi/2$ and (a): $m=1$, (b): $m=3$. See Sec. \ref{NonZeroEfield_phiE_m_Study}, item 2 for discussion.
}
\label{Fig4}
\end{figure}

\item $\phi_E=\pi/4$ or $3\pi/4$.\\
For these tilted electric field directions the ordered stripe pattern emerges only at [100] surface orientation, $m=2$, and the stripes also make the field-imposed angle
with the $x$-axis, see Fig. \ref{Fig5}(a-c). For other surface orientations the stripe pattern is not perfect at computed times. For instance, in Fig. \ref{Fig5}(d) one can notice 
zipped stripes, and the angle that these stripes
make with the $x$ axis may significantly deviate from $\phi_E$  -  more so as computed evolution gets longer. Fig. \ref{Fig5}(a-c) shows the time-progression of the 
coarsening of the perfect stripe pattern, i.e. 
the reduction of the aerial number density of the stripes with a high $C_B$ content (red stripes). 
Note that the number of red stripes is reduced from 14 in Fig. \ref{Fig5}(a) to 8 in Fig. \ref{Fig5}(c). Fig. \ref{Fig5}(b) captures the 
moment when some red stripes are dissolving, which increases the width of low $C_B$ content purple stripes.\\ 

\emph{Remark 5.}\; Some coarsening is present for all cases computed in this paper, but we stopped short of quantitatively characterizing this process, 
since this requires computing on very long time scales. We presently do not possess the computational resources for this task. It is important to underscore that a perfect 
stripe pattern always emerges at the early stages in the evolution and it stays perfect regardless of the coarsening rate. However, a multiple-connected domains 
pattern never transforms into a perfect stripe pattern at later times, and a zipped stripe pattern, such as one in Fig. \ref{Fig5}(d), infrequently becomes a perfect 
stripe pattern at very large times.
\begin{figure}[H]
\vspace{-0.2cm}
\centering
\includegraphics[width=5.0in]{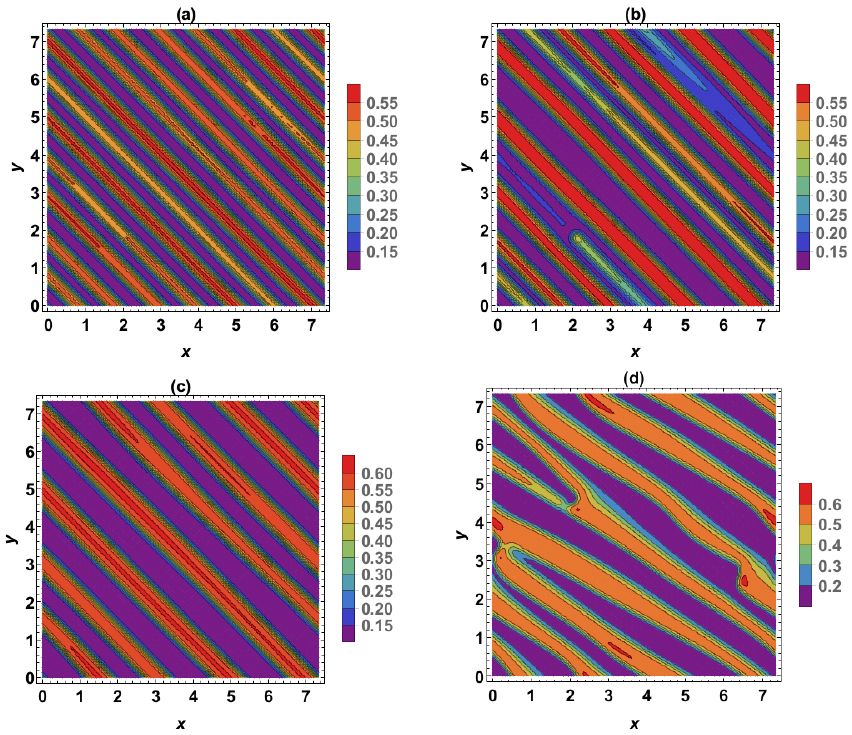}
\vspace{-0.15cm}
\caption{Examples of composition patterns. (a-c): $C_B(x,y,300),\ C_B(x,y,800),\ C_B(x,y,1100)$ at $\phi_E=3\pi/4$ and $m=2$. (d): $C_B(x,y,300)$ at $\phi_E=3\pi/4$ and $m=1$. 
See Sec. \ref{NonZeroEfield_phiE_m_Study}, item 3 for discussion.
}
\label{Fig5}
\end{figure}

\end{enumerate} 

\subsubsection{Zero crystallographic anisotropy, non-zero diffusional anisotropy: $\beta=0$}
\label{NonZeroEfield_phiE_Lambda_Study}

In this subsection the crystallographic anisotropy is zero, thus values of $m$ and $\phi^{(x)}$ are irrelevant. We summarize the patterns at various direction angles of 
the electric field, $\phi_E$, and at various diffusional anisotropies $\Lambda_B$. 

\begin{enumerate} 

\item $\phi_E=0$ or $\pi$.\\
Horizontal stripe pattern at $\Lambda_B\le 1$, and multiple-connected domains at $\Lambda_B>1$. For $\Lambda_B\le 1$ the aerial number density decreases as $\Lambda_B$ increases.

\item $\phi_E=\pi/2$ or $-\pi/2$.\\
Disordered pattern at $\Lambda_B< 1$ (similar to Fig. \ref{Fig2}(b)); perfect vertical stripes at $\Lambda_B\ge 1$. The aerial number density depends on
$\Lambda_B$ non-monotonously, first decreasing at $1\le \Lambda_B\le 2$ and then increasing at $2< \Lambda_B\le 6$ and reaching the maximum value of 7 or 8 (Fig. \ref{Fig6}(a)).
The times of the stripe pattern formation, $t_{stripe}$, shown in Fig. \ref{Fig6}(b) decrease rapidly as $\Lambda_B$ increases, reach the minimum value 1000 at $\Lambda_B=6$, and stay constant as $\Lambda_B$ 
further increases. 
%
\begin{figure}[H]
\vspace{-0.2cm}
\centering
\includegraphics[width=5.0in]{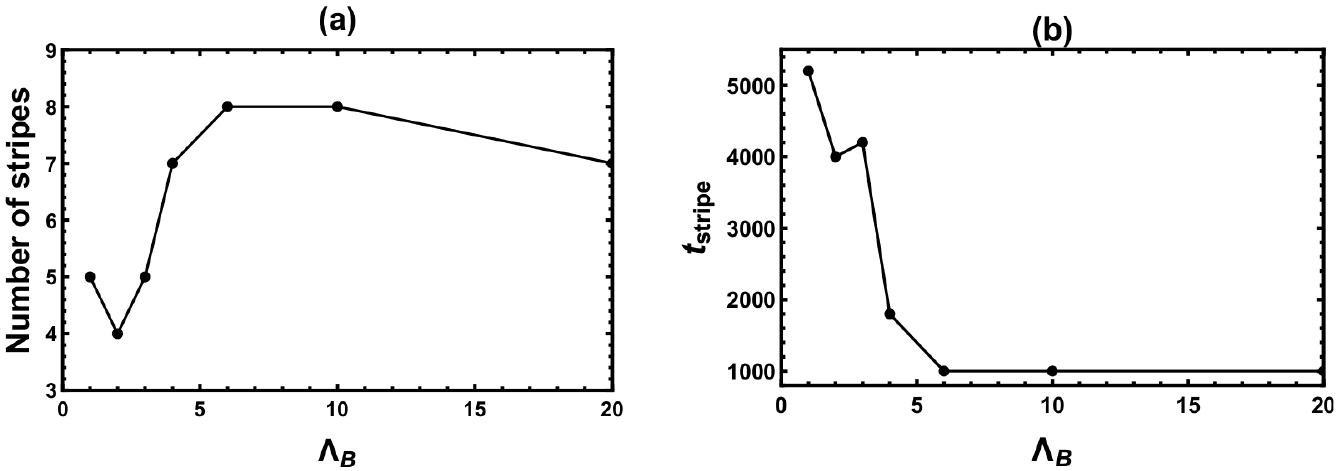}
\vspace{-0.15cm}
\caption{(a): The number of vertical stripes vs. $\Lambda_B$. For each $\Lambda_B$ value shown, the stripes are counted when the stripe pattern first emerges. 
These times are plotted in (b) vs. $\Lambda_B$.   
See Sec. \ref{NonZeroEfield_phiE_Lambda_Study}, item 2 for discussion. 
}
\label{Fig6}
\end{figure}

\item $\phi_E=\pi/4$ or $3\pi/4$.\\
Zipped stripe pattern at $\Lambda_B<1$, transforming into a perfect stripe pattern at $\Lambda_B\ge 1$. The aerial number density increases as $\Lambda_B$ increases for
$\phi_E=\pi/4$, and decreases as $\Lambda_B$ increases for $\phi_E=3\pi/4$. Stripes become more vertical as $\Lambda_B$
increases, see Fig. \ref{Fig7} (thus the pattern orientation angle deviates from $\phi_E$ for $\Lambda_B> 1$).
\begin{figure}[H]
\vspace{-0.2cm}
\centering
\includegraphics[width=5.0in]{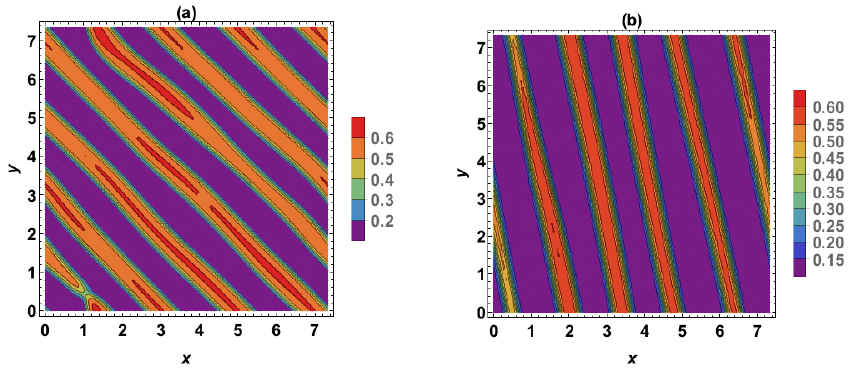}
\vspace{-0.15cm}
\caption{Examples of composition patterns. $C_B(x,y,2200)$ at $\phi_E=3\pi/4$ and (a): $\Lambda_B=1$, (b): $\Lambda_B=6$. See Sec. \ref{NonZeroEfield_phiE_Lambda_Study}, item 3 for discussion.
}
\label{Fig7}
\end{figure}

\end{enumerate}

\subsubsection{Crystallographic anisotropy and diffusional anisotropy both non-zero: $m=1,\ \beta=5,\ \Lambda_B=2,\ \phi_E=\pi/4$}
\label{NonZeroEfield_FullAnizCase_phiX_Study}

In this subsection we discuss the computation of the composition evolution in [110]-oriented film with the full anisotropy effect. Such situation is most relevant to experiment.

\begin{figure}[H]
\vspace{-0.2cm}
\centering
\includegraphics[width=7.0in]{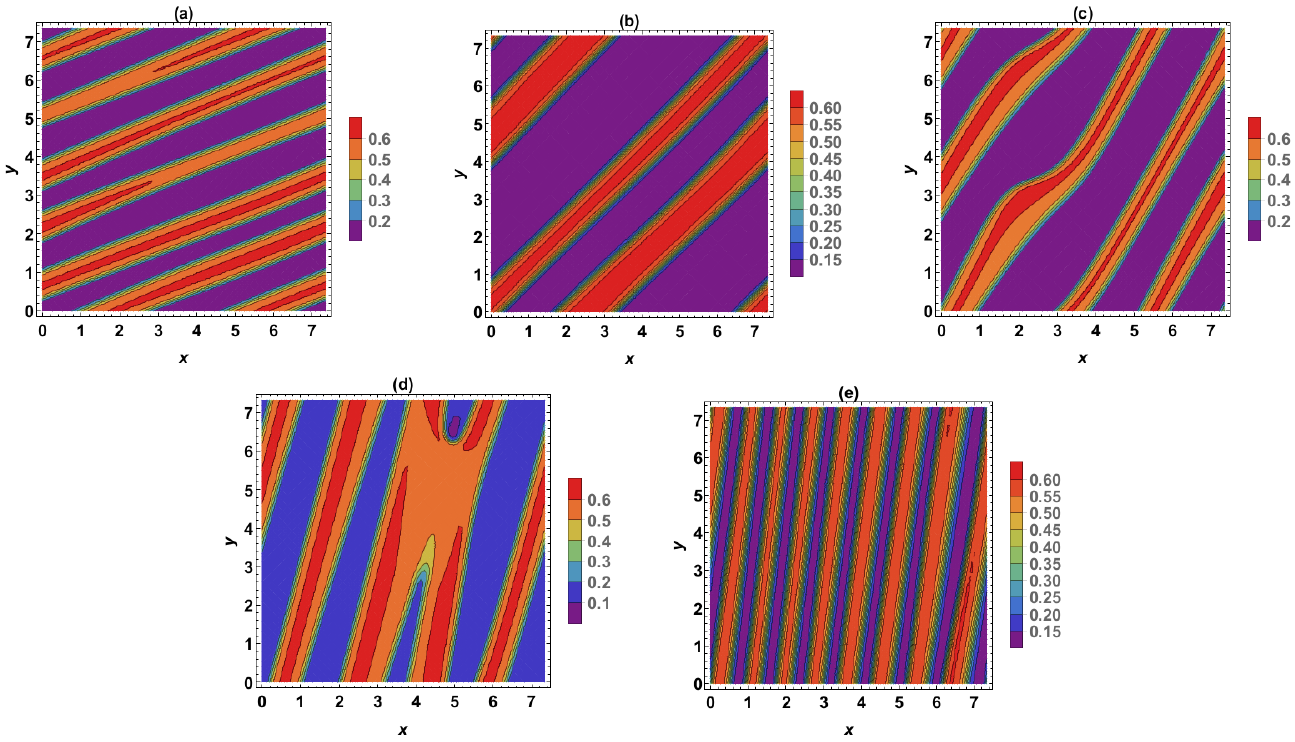}
\vspace{-0.15cm}
\caption{Examples of composition patterns. (a): $t=2000,\ \phi^{(x)}=0$, (b): $t=3000,\ \phi^{(x)}=\pi/6$, (c): $t=2000,\ \phi^{(x)}=\pi/4$, (d): $t=4000,\ \phi^{(x)}=\pi/3$, 
(e): $t=3000,\ \phi^{(x)}=\pi/2$. The stripe patterns in (a)-(c) and (e) are shown after they just formed, i.e. there has been no coarsening yet. See Sec. \ref{NonZeroEfield_FullAnizCase_phiX_Study} for discussion.
}
\label{Fig8}
\end{figure}

Fig. \ref{Fig8} shows the impact of the diffusion misorientation angle $\phi^{(x)}$. The competition of the electric field 
and the anisotropy is very striking. The electric field tries to align the stripe pattern at $45^\circ$ angle to the $x$ axis, but the diffusion anisotropy interferes, and only at
$\phi^{(x)}=\pi/6$ the $45^\circ$ angle is achieved. 
It may be said that at $\phi^{(x)}=\pi/6$ the anisotropy and the electric field are aligned, creating the ``constructive interference effect".
Conversely, at $\phi^{(x)}=\pi/3$, unlike at other $\phi^{(x)}$ values, the stripe pattern is double-zipped - the misalignment between the electric field 
and the anisotropy creates the ``destructive interference effect". (Notice also that at $\phi^{(x)}=\pi/4$ the stripes are present, but they are not straight and their 
inclination angle to the $x$-axis is larger than $45^\circ$.)
Between these two extremes, the stripe inclination angle  reflects the balance between 
$\phi^{(x)}$ and $\phi_E$, and the aerial number density of the stripes is also affected by how close are these angles. It is the largest at $\phi^{(x)}=0$ and $\pi/2$, 
i.e. when $\phi^{(x)}$ deviates the most from $\phi_E$. 

It is important to emphasize here that also for [100] and [111]-oriented films there does exist a combination of the parameters ensuring that the stripes form in
the electric field direction. For instance, in [111] case, keeping $\beta=5,\ \Lambda_B=2,\ \phi_E=\pi/4$, the stripes form at $45^\circ$ angle to the $x$ axis provided that
$\phi^{(x)}=\pi/16$.

\begin{figure}[H]
\vspace{-0.2cm}
\centering
\includegraphics[width=7.0in]{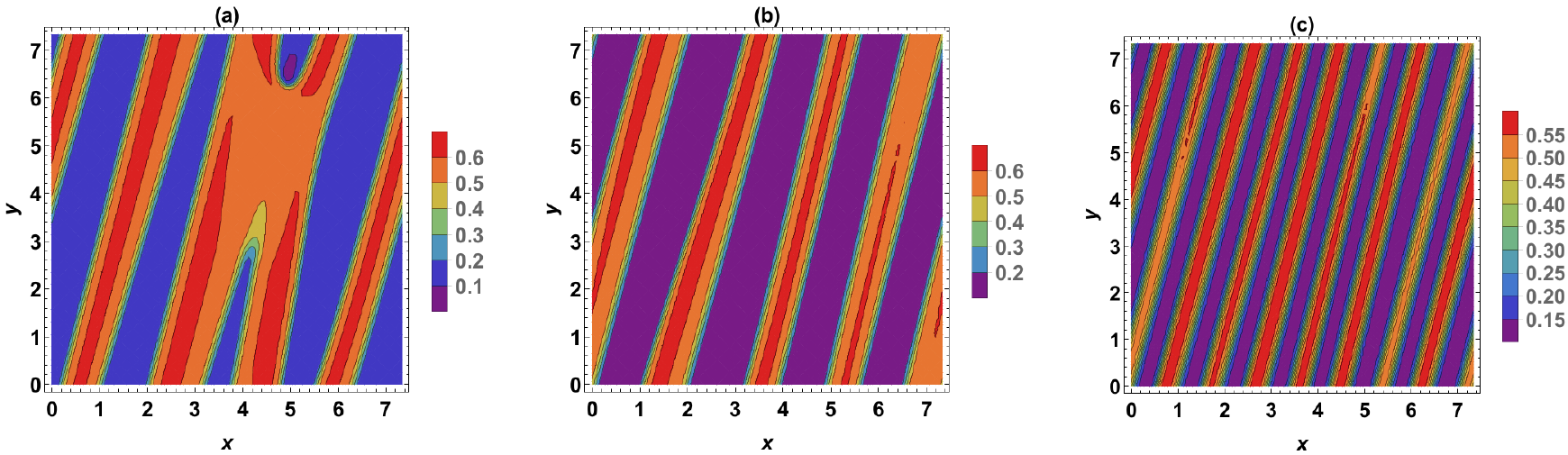}
\vspace{-0.15cm}
\caption{Examples of composition patterns. (a): $t=4000,\ \Delta V=0.03$V, (b): $t=1400,\ \Delta V=0.1$V, (c): $t=400,\ \Delta V=0.15$V. Panel (a) is the copy of Fig. \ref{Fig8}(d).
The stripe patterns in (b) and (c) are shown after they just formed, i.e. there has been no coarsening yet. See Sec. \ref{NonZeroEfield_FullAnizCase_phiX_Study} for discussion.
}
\label{Fig9}
\end{figure}

Lastly, Fig. \ref{Fig9} shows the impact of the last parameter that remains to study - the applied voltage $\Delta V$. We chose all parameters as in Fig. \ref{Fig8}(d) and increased $\Delta V$.
It is obvious that at higher voltage the quality of the stripe pattern is improved. Another effect is the increased aerial number density. However, the alignment to the electric
field direction, $\phi_E=\pi/4$, does not improve when the electromigration effect gets stronger. Increasing $\Delta V$ beyond $0.15$V we did not detect any further changes to the pattern that is shown in 
Fig. \ref{Fig9}(c). It is worth observing that these results transparently show that the aerial number density (equivalently, the wavelength of the 1D spatially periodic stripe pattern) 
is unrelated to the LSA most dangerous wavelength $\lambda_{max}$\footnote{$\lambda_{max}=2\pi/k_{max}$, where $k_{max}=\mbox{max}\left(k_{1max},k_{2max}\right)$, 
and $k_{1max},\ k_{2max}$  maximize $\omega_r\left(k_1,k_2\right)$.}, 
since the latter wavelength stems from the real part of the perturbation growth rate, whereas $\Delta V$ enters only the 
imaginary part of that rate. Thus the aerial number density should be determined by the nonlinear interaction of the lateral concentration waves.
%

Based on the computations in this section, for the most practical [110] surface orientation with the mild total anisotropy, $0<\beta,\Lambda <10$, 
and at a fixed electric field strength the stripes 
angle $\phi_{st}$ satisfies 
$|\phi_{st}-\phi_E|\le \kappa\left(\phi^{(x)}\right)$,
where $\kappa\left(\phi^{(x)}\right)$ is the positive function that attains a minimum on its domain $\phi_E-u_1\le \phi^{(x)}\le \phi_E+u_2$. Here $u_1$ and $u_2$ are the 
positive angles,
such that $\phi^{(x)}$ is in the same quadrant with $\phi_E$. The aerial number density of the stripes is the largest at the endpoints $\phi_E-u_1$ and $\phi_E+u_2$.
Increasing the strength of the applied electric field below a certain upper limit (while keeping $\phi_E$ and $\phi^{(x)}$ fixed) eliminates the imperfect stripes, increases the aerial 
number density of the perfect ones, and shortens the formation time of a perfect stripe pattern.

\section{Summary}
\label{Conc}

We demonstrated how the complex interaction of the diffusion and electromigration 
influences the formation and evolution of the stripe composition patterns in a thin, crystallographically oriented surface layer 
of a heated substitutional binary alloy film. 
With a full surface diffusion anisotropy, there is no single control parameter that is responsible for both 
the stripes alignment to the electric field direction and for the aerial number density of the stripes. Tuning the applied voltage allows to 
adjust the density, and tuning the electric field direction allows to form the stripes that are closely aligned to that direction. 
If the anisotropy parameters are certain, then the computer simulation of this model would inform the experiment about the expected pattern.





\end{document}